
%
%
%

\catcode `\@=11 

\def\@version{1.3}
\def\@verdate{28.11.1992}

\font\fiverm=cmr5
\font\fivei=cmmi5	\skewchar\fivei='177
\font\fivesy=cmsy5	\skewchar\fivesy='60
\font\fivebf=cmbx5

\font\sevenrm=cmr7
\font\seveni=cmmi7	\skewchar\seveni='177
\font\sevensy=cmsy7	\skewchar\sevensy='60
\font\sevenbf=cmbx7

\font\eightrm=cmr8
\font\eightbf=cmbx8
\font\eightit=cmti8
\font\eighti=cmmi8			\skewchar\eighti='177
\font\eightmib=cmmib10 at 8pt	\skewchar\eightmib='177
\font\eightsy=cmsy8			\skewchar\eightsy='60
\font\eightsyb=cmbsy10 at 8pt	\skewchar\eightsyb='60
\font\eightsl=cmsl8
\font\eighttt=cmtt8			\hyphenchar\eighttt=-1
\font\eightcsc=cmcsc10 at 8pt
\font\eightsf=cmss8

\font\ninerm=cmr9
\font\ninebf=cmbx9
\font\nineit=cmti9
\font\ninei=cmmi9			\skewchar\ninei='177
\font\ninemib=cmmib10 at 9pt	\skewchar\ninemib='177
\font\ninesy=cmsy9			\skewchar\ninesy='60
\font\ninesyb=cmbsy10 at 9pt	\skewchar\ninesyb='60
\font\ninesl=cmsl9
\font\ninett=cmtt9			\hyphenchar\ninett=-1
\font\ninecsc=cmcsc10 at 9pt
\font\ninesf=cmss9

\font\tenrm=cmr10
\font\tenbf=cmbx10
\font\tenit=cmti10
\font\teni=cmmi10		\skewchar\teni='177
\font\tenmib=cmmib10	\skewchar\tenmib='177
\font\tensy=cmsy10		\skewchar\tensy='60
\font\tensyb=cmbsy10	\skewchar\tensyb='60
\font\tenex=cmex10
\font\tensl=cmsl10
\font\tentt=cmtt10		\hyphenchar\tentt=-1
\font\tencsc=cmcsc10
\font\tensf=cmss10

\font\elevenrm=cmr10 scaled \magstephalf
\font\elevenbf=cmbx10 scaled \magstephalf
\font\elevenit=cmti10 scaled \magstephalf
\font\eleveni=cmmi10 scaled \magstephalf	\skewchar\eleveni='177
\font\elevenmib=cmmib10 scaled \magstephalf	\skewchar\elevenmib='177
\font\elevensy=cmsy10 scaled \magstephalf	\skewchar\elevensy='60
\font\elevensyb=cmbsy10 scaled \magstephalf	\skewchar\elevensyb='60
\font\elevensl=cmsl10 scaled \magstephalf
\font\eleventt=cmtt10 scaled \magstephalf	\hyphenchar\eleventt=-1
\font\elevencsc=cmcsc10 scaled \magstephalf
\font\elevensf=cmss10 scaled \magstephalf

\font\fourteenrm=cmr10 scaled \magstep2
\font\fourteenbf=cmbx10 scaled \magstep2
\font\fourteenit=cmti10 scaled \magstep2
\font\fourteeni=cmmi10 scaled \magstep2		\skewchar\fourteeni='177
\font\fourteenmib=cmmib10 scaled \magstep2	\skewchar\fourteenmib='177
\font\fourteensy=cmsy10 scaled \magstep2	\skewchar\fourteensy='60
\font\fourteensyb=cmbsy10 scaled \magstep2	\skewchar\fourteensyb='60
\font\fourteensl=cmsl10 scaled \magstep2
\font\fourteentt=cmtt10 scaled \magstep2	\hyphenchar\fourteentt=-1
\font\fourteencsc=cmcsc10 scaled \magstep2
\font\fourteensf=cmss10 scaled \magstep2

\font\seventeenrm=cmr10 scaled \magstep3
\font\seventeenbf=cmbx10 scaled \magstep3
\font\seventeenit=cmti10 scaled \magstep3
\font\seventeeni=cmmi10 scaled \magstep3	\skewchar\seventeeni='177
\font\seventeenmib=cmmib10 scaled \magstep3	\skewchar\seventeenmib='177
\font\seventeensy=cmsy10 scaled \magstep3	\skewchar\seventeensy='60
\font\seventeensyb=cmbsy10 scaled \magstep3	\skewchar\seventeensyb='60
\font\seventeensl=cmsl10 scaled \magstep3
\font\seventeentt=cmtt10 scaled \magstep3	\hyphenchar\seventeentt=-1
\font\seventeencsc=cmcsc10 scaled \magstep3
\font\seventeensf=cmss10 scaled \magstep3

\def\@typeface{Computer Modern} 

\def\hexnumber@#1{\ifnum#1<10 \number#1\else
 \ifnum#1=10 A\else\ifnum#1=11 B\else\ifnum#1=12 C\else
 \ifnum#1=13 D\else\ifnum#1=14 E\else\ifnum#1=15 F\fi\fi\fi\fi\fi\fi\fi}

\def\mib{\hexnumber@\mibfam}
\def\syb{\hexnumber@\sybfam}

\def\makestrut{%
  \setbox\strutbox=\hbox{%
    \vrule height.7\baselineskip depth.3\baselineskip width 0pt}%
}

\def\bls#1{%
  \normalbaselineskip=#1%
  \normalbaselines%
  \makestrut%
}

%

\newfam\mibfam 
\newfam\sybfam 
\newfam\scfam  
\newfam\sffam  

\def\em{\ifdim\fontdimen1\font>0 \rm\else\it\fi}

\textfont3=\tenex
\scriptfont3=\tenex
\scriptscriptfont3=\tenex

\def\eightpoint{
  \def\rm{\fam0\eightrm}%
  \textfont0=\eightrm \scriptfont0=\sevenrm \scriptscriptfont0=\fiverm%
  \textfont1=\eighti  \scriptfont1=\seveni  \scriptscriptfont1=\fivei%
  \textfont2=\eightsy \scriptfont2=\sevensy \scriptscriptfont2=\fivesy%
  \textfont\itfam=\eightit\def\it{\fam\itfam\eightit}%
  \textfont\bffam=\eightbf%
    \scriptfont\bffam=\sevenbf%
      \scriptscriptfont\bffam=\fivebf%
  \def\bf{\fam\bffam\eightbf}%
  \textfont\slfam=\eightsl\def\sl{\fam\slfam\eightsl}%
  \textfont\ttfam=\eighttt\def\tt{\fam\ttfam\eighttt}%
  \textfont\scfam=\eightcsc\def\sc{\fam\scfam\eightcsc}%
  \textfont\sffam=\eightsf\def\sf{\fam\sffam\eightsf}%
  \textfont\mibfam=\eightmib%
  \textfont\sybfam=\eightsyb%
  \bls{10pt}%
}

\def\ninepoint{
  \def\rm{\fam0\ninerm}%
  \textfont0=\ninerm \scriptfont0=\sevenrm \scriptscriptfont0=\fiverm%
  \textfont1=\ninei  \scriptfont1=\seveni  \scriptscriptfont1=\fivei%
  \textfont2=\ninesy \scriptfont2=\sevensy \scriptscriptfont2=\fivesy%
  \textfont\itfam=\nineit\def\it{\fam\itfam\nineit}%
  \textfont\bffam=\ninebf%
    \scriptfont\bffam=\sevenbf%
      \scriptscriptfont\bffam=\fivebf%
  \def\bf{\fam\bffam\ninebf}%
  \textfont\slfam=\ninesl\def\sl{\fam\slfam\ninesl}%
  \textfont\ttfam=\ninett\def\tt{\fam\ttfam\ninett}%
  \textfont\scfam=\ninecsc\def\sc{\fam\scfam\ninecsc}%
  \textfont\sffam=\ninesf\def\sf{\fam\sffam\ninesf}%
  \textfont\mibfam=\ninemib%
  \textfont\sybfam=\ninesyb%
  \bls{12pt}%
}

\def\tenpoint{
  \def\rm{\fam0\tenrm}%
  \textfont0=\tenrm \scriptfont0=\sevenrm \scriptscriptfont0=\fiverm%
  \textfont1=\teni  \scriptfont1=\seveni  \scriptscriptfont1=\fivei%
  \textfont2=\tensy \scriptfont2=\sevensy \scriptscriptfont2=\fivesy%
  \textfont\itfam=\tenit\def\it{\fam\itfam\tenit}%
  \textfont\bffam=\tenbf%
    \scriptfont\bffam=\sevenbf%
      \scriptscriptfont\bffam=\fivebf%
  \def\bf{\fam\bffam\tenbf}%
  \textfont\slfam=\tensl\def\sl{\fam\slfam\tensl}%
  \textfont\ttfam=\tentt\def\tt{\fam\ttfam\tentt}%
  \textfont\scfam=\tencsc\def\sc{\fam\scfam\tencsc}%
  \textfont\sffam=\tensf\def\sf{\fam\sffam\tensf}%
  \textfont\mibfam=\tenmib%
  \textfont\sybfam=\tensyb%
  \bls{12pt}%
}

\def\elevenpoint{
  \def\rm{\fam0\elevenrm}%
  \textfont0=\elevenrm \scriptfont0=\eightrm \scriptscriptfont0=\fiverm%
  \textfont1=\eleveni  \scriptfont1=\eighti  \scriptscriptfont1=\fivei%
  \textfont2=\elevensy \scriptfont2=\eightsy \scriptscriptfont2=\fivesy%
  \textfont\itfam=\elevenit\def\it{\fam\itfam\elevenit}%
  \textfont\bffam=\elevenbf%
    \scriptfont\bffam=\eightbf%
      \scriptscriptfont\bffam=\fivebf%
  \def\bf{\fam\bffam\elevenbf}%
  \textfont\slfam=\elevensl\def\sl{\fam\slfam\elevensl}%
  \textfont\ttfam=\eleventt\def\tt{\fam\ttfam\eleventt}%
  \textfont\scfam=\elevencsc\def\sc{\fam\scfam\elevencsc}%
  \textfont\sffam=\elevensf\def\sf{\fam\sffam\elevensf}%
  \textfont\mibfam=\elevenmib%
  \textfont\sybfam=\elevensyb%
  \bls{13pt}%
}

\def\fourteenpoint{
  \def\rm{\fam0\fourteenrm}%
  \textfont0\fourteenrm  \scriptfont0\tenrm  \scriptscriptfont0\sevenrm%
  \textfont1\fourteeni   \scriptfont1\teni   \scriptscriptfont1\seveni%
  \textfont2\fourteensy  \scriptfont2\tensy  \scriptscriptfont2\sevensy%
  \textfont\itfam=\fourteenit\def\it{\fam\itfam\fourteenit}%
  \textfont\bffam=\fourteenbf%
    \scriptfont\bffam=\tenbf%
      \scriptscriptfont\bffam=\sevenbf%
  \def\bf{\fam\bffam\fourteenbf}%
  \textfont\slfam=\fourteensl\def\sl{\fam\slfam\fourteensl}%
  \textfont\ttfam=\fourteentt\def\tt{\fam\ttfam\fourteentt}%
  \textfont\scfam=\fourteencsc\def\sc{\fam\scfam\fourteencsc}%
  \textfont\sffam=\fourteensf\def\sf{\fam\sffam\fourteensf}%
  \textfont\mibfam=\fourteenmib%
  \textfont\sybfam=\fourteensyb%
  \bls{17pt}%
}

\def\seventeenpoint{
  \def\rm{\fam0\seventeenrm}%
  \textfont0\seventeenrm  \scriptfont0\elevenrm  \scriptscriptfont0\ninerm%
  \textfont1\seventeeni   \scriptfont1\eleveni   \scriptscriptfont1\ninei%
  \textfont2\seventeensy  \scriptfont2\elevensy  \scriptscriptfont2\ninesy%
  \textfont\itfam=\seventeenit\def\it{\fam\itfam\seventeenit}%
  \textfont\bffam=\seventeenbf%
    \scriptfont\bffam=\elevenbf%
      \scriptscriptfont\bffam=\ninebf%
  \def\bf{\fam\bffam\seventeenbf}%
  \textfont\slfam=\seventeensl\def\sl{\fam\slfam\seventeensl}%
  \textfont\ttfam=\seventeentt\def\tt{\fam\ttfam\seventeentt}%
  \textfont\scfam=\seventeencsc\def\sc{\fam\scfam\seventeencsc}%
  \textfont\sffam=\seventeensf\def\sf{\fam\sffam\seventeensf}%
  \textfont\mibfam=\seventeenmib%
  \textfont\sybfam=\seventeensyb%
  \bls{20pt}%
}

\lineskip=1pt      \normallineskip=\lineskip
\lineskiplimit=0pt \normallineskiplimit=\lineskiplimit




\def\Nulle{0}  
\def\Aue{1}    
\def\Afe{2}    
\def\Sue{4}    
\def\Hae{5}    
\def\Hbe{6}    
\def\Hce{7}    
\def\Hde{8}    
\def\Kwe{9}    
\def\Txe{10}   
\def\Lie{11}   
\def\Bbe{12}   


\newdimen\DimenA
\newbox\BoxA

\newcount\LastMac \LastMac=\Nulle
\newcount\HeaderNumber \HeaderNumber=0
\newcount\DefaultHeader \DefaultHeader=\HeaderNumber
\newskip\Indent

\newskip\half      \half=5.5pt plus 1.5pt minus 2.25pt
\newskip\one       \one=11pt plus 3pt minus 5.5pt
\newskip\onehalf   \onehalf=16.5pt plus 5.5pt minus 8.25pt
\newskip\two       \two=22pt plus 5.5pt minus 11pt

\def\Half{\vskip-\lastskip\vskip\half}
\def\One{\vskip-\lastskip\vskip\one}
\def\OneHalf{\vskip-\lastskip\vskip\onehalf}
\def\Two{\vskip-\lastskip\vskip\two}


\def\rTenPT{10pt plus \Feathering}

\def\TenPT{10pt plus \Feathering} 
\def\ElevenPT{11pt plus \Feathering}

\def\Raggedright{
 \rightskip=0pt plus \hsize
}

\def\Fullout{
\rightskip=0pt
}

\def\Hang#1#2{
 \hangindent=#1
 \hangafter=#2
}

\def\EveryMac{
 \Fullout
 \everypar{}
}



\def\title#1{
 \EveryMac
 \LastMac=\Nulle
 \global\HeaderNumber=0
 \global\DefaultHeader=1
 \vbox to 1pc{\vss}
 \seventeenpoint
 \Raggedright
 \noindent \bf #1
}

\def\author#1{
 \EveryMac
 \ifnum\LastMac=\Afe \OneHalf
  \else \Two
 \fi
 \LastMac=\Aue
 \fourteenpoint
 \Raggedright
 \noindent \rm #1\par
 \vskip 3pt\relax
}

\def\affiliation#1{
 \EveryMac
 \LastMac=\Afe
 \eightpoint\bls{\TenPT}
 \Raggedright
 \noindent \it #1\par
}

\def\abstract{%
 \EveryMac
 \Two
 \LastMac=\Sue
 \everypar{\Hang{11pc}{0}}
 \noindent\ninebf ABSTRACT\par
 \tenpoint\bls{\ElevenPT}
 \Fullout
 \noindent\rm
}

\def\keywords{
 \EveryMac
 \Half
 \LastMac=\Kwe
 \everypar{\Hang{11pc}{0}}
 \tenpoint\bls{\ElevenPT}
 \Fullout
 \noindent\hbox{\bf Key words:\ }
 \rm
}


\def\maketitle{%
  \Two%
  \EndOpening%
  \MakePage%
}



\def\Autonumber{
 \global\AutoNumbertrue  
}

\newif\ifAutoNumber \AutoNumberfalse
\newcount\Sec        
\newcount\SecSec
\newcount\SecSecSec

\Sec=0

\def\:{\let\@sptoken= } \:  
\def\:{\@xifnch} \expandafter\def\: {\futurelet\@tempc\@ifnch}

\def\@ifnextchar#1#2#3{%
  \let\@tempMACe #1%
  \def\@tempMACa{#2}%
  \def\@tempMACb{#3}%
  \futurelet \@tempMACc\@ifnch%
}

\def\@ifnch{%
\ifx \@tempMACc \@sptoken%
  \let\@tempMACd\@xifnch%
\else%
  \ifx \@tempMACc \@tempMACe%
    \let\@tempMACd\@tempMACa%
  \else%
    \let\@tempMACd\@tempMACb%
  \fi%
\fi%
\@tempMACd%
}

\def\@ifstar#1#2{\@ifnextchar *{\def\@tempMACa*{#1}\@tempMACa}{#2}}

\def\section{\@ifstar{\@ssection}{\@section}}

\def\@section#1{
 \EveryMac
 \Two
 \LastMac=\Hae
 \ninepoint\bls{\ElevenPT}
 \bf
 \Raggedright
 \ifAutoNumber
  \advance\Sec by 1
  \noindent\number\Sec\hskip 1pc \uppercase{#1}
  \SecSec=0
 \else
  \noindent \uppercase{#1}
 \fi
 \nobreak
}

\def\@ssection#1{
 \EveryMac
 \ifnum\LastMac=\Hae \Half
  \else \OneHalf
 \fi
 \LastMac=\Hae
 \tenpoint\bls{\ElevenPT}
 \bf
 \Raggedright
 \noindent\uppercase{#1}
}

\def\subsection#1{
 \EveryMac
 \ifnum\LastMac=\Hae \Half
  \else \OneHalf
 \fi
 \LastMac=\Hbe
 \tenpoint\bls{\ElevenPT}
 \bf
 \Raggedright
 \ifAutoNumber
  \advance\SecSec by 1
  \noindent\number\Sec.\number\SecSec
  \hskip 1pc #1
  \SecSecSec=0
 \else
  \noindent #1
 \fi
 \nobreak
}

\def\subsubsection#1{
 \EveryMac
 \ifnum\LastMac=\Hbe \Half
  \else \OneHalf
 \fi
 \LastMac=\Hce
 \ninepoint\bls{\ElevenPT}
 \it
 \Raggedright
 \ifAutoNumber
  \advance\SecSecSec by 1
  \noindent\number\Sec.\number\SecSec.\number\SecSecSec
  \hskip 1pc #1
 \else
  \noindent #1
 \fi
 \nobreak
}

\def\paragraph#1{
 \EveryMac
 \One
 \LastMac=\Hde
 \ninepoint\bls{\ElevenPT}
 \noindent \it #1
 \rm
}


\def\tx{
 \EveryMac
 \ifnum\LastMac=\Lie \Half\fi
 \ifnum\LastMac=\Hae \nobreak\Half\fi
 \ifnum\LastMac=\Hbe \nobreak\Half\fi
 \ifnum\LastMac=\Hce \nobreak\Half\fi
 \ifnum\LastMac=\Lie \else \noindent\fi
 \LastMac=\Txe
 \ninepoint\bls{\ElevenPT}
 \rm
}


\def\item{
 \par
 \EveryMac
 \ifnum\LastMac=\Lie
  \else \Half
 \fi
 \LastMac=\Lie
 \ninepoint\bls{\ElevenPT}
 \rm
}


\def\bibitem{
 \par
 \EveryMac
 \ifnum\LastMac=\Bbe
  \else \Half
 \fi
 \LastMac=\Bbe
 \Hang{1.5em}{1}
 \eightpoint\bls{\TenPT}
 \Raggedright
 \noindent \rm
}


\newtoks\CatchLine

\def\@journal{Mon.\ Not.\ R.\ Astron.\ Soc.\ }  
\def\@pubyear{1993}        
\def\@pagerange{000--000}  
\def\@volume{000}          
\def\@microfiche{}         %

\def\pubyear#1{\gdef\@pubyear{#1}\@makecatchline}
\def\pagerange#1{\gdef\@pagerange{#1}\@makecatchline}
\def\volume#1{\gdef\@volume{#1}\@makecatchline}
\def\microfiche#1{\gdef\@microfiche{and Microfiche\ #1}\@makecatchline}

\def\@makecatchline{%
  \global\CatchLine{%
    {\rm \@journal {\bf \@volume},\ \@pagerange\ (\@pubyear)\ \@microfiche}}%
}

\@makecatchline 

\newtoks\LeftHeader
\def\shortauthor#1{
 \global\LeftHeader{#1}
}

\newtoks\RightHeader
\def\shorttitle#1{
 \global\RightHeader{#1}
}

\def\PageHead{
 \EveryMac
 \ifnum\HeaderNumber=1 \Pagehead
  \else \Catchline
 \fi
}

\def\Catchline{%
 \vbox to 0pt{\vskip-22.5pt
  \hbox to \PageWidth{\vbox to8.5pt{}\noindent
  \eightpoint\the\CatchLine\hfill}\vss}
 \nointerlineskip
}

\def\Pagehead{%
 \ifodd\pageno
   \vbox to 0pt{\vskip-22.5pt
   \hbox to \PageWidth{\vbox to8.5pt{}\elevenpoint\it\noindent
    \hfill\the\RightHeader\hskip1.5em\rm\folio}\vss}
 \else
   \vbox to 0pt{\vskip-22.5pt
   \hbox to \PageWidth{\vbox to8.5pt{}\elevenpoint\rm\noindent
   \folio\hskip1.5em\it\the\LeftHeader\hfill}\vss}
 \fi
 \nointerlineskip
}

\def\PageFoot{} 

\def\authorcomment#1{%
  \gdef\PageFoot{%
    \nointerlineskip%
    \vbox to 22pt{\vfil%
      \hbox to \PageWidth{\elevenpoint\rm\noindent \hfil #1 \hfil}}%
  }%
}

\everydisplay{\displaysetup}

\newif\ifeqno
\newif\ifleqno

\def\displaysetup#1$${%
 \displaytest#1\eqno\eqno\displaytest
}

\def\displaytest#1\eqno#2\eqno#3\displaytest{%
 \if!#3!\ldisplaytest#1\leqno\leqno\ldisplaytest
 \else\eqnotrue\leqnofalse\def\eqn{#2}\def\eq{#1}\fi
 \generaldisplay$$}

\def\ldisplaytest#1\leqno#2\leqno#3\ldisplaytest{%
 \def\eq{#1}%
 \if!#3!\eqnofalse\else\eqnotrue\leqnotrue
  \def\eqn{#2}\fi}

\def\generaldisplay{%
\ifeqno \ifleqno 
   \hbox to \hsize{\noindent
     $\displaystyle\eq$\hfil$\displaystyle\eqn$}
  \else
    \hbox to \hsize{\noindent
     $\displaystyle\eq$\hfil$\displaystyle\eqn$}
  \fi
 \else
 \hbox to \hsize{\vbox{\noindent
  $\displaystyle\eq$\hfil}}
 \fi
}

\def\@notice{%
  \par\Two%
  \bls{12pt}%
  \noindent\tenrm This paper has been produced using the Blackwell
                  Scientific Publications \TeX\ macros.%
}

\outer\def\bye{\@notice\par\vfill\supereject\end}

\everyjob{%
  \Warn{Monthly notices of the RAS journal style (\@typeface)\space
        v\@version,\space \@verdate.}\Warn{}%
}




\newif\if@debug \@debugfalse  

\def\Print#1{\if@debug\immediate\write16{#1}\else \fi}
\def\Warn#1{\immediate\write16{#1}}
\def\wlog#1{}

\newcount\Iteration 

\newif\ifFigureBoxes  
\FigureBoxestrue

\def\Single{0} \def\Double{1}                 
\def\Figure{0} \def\Table{1}                  

\def\InStack{0}  
\def\InZoneA{1}
\def\InZoneB{2}
\def\InZoneC{3}

\newcount\TEMPCOUNT 
\newdimen\TEMPDIMEN 
\newbox\TEMPBOX     
\newbox\VOIDBOX     

\newcount\LengthOfStack 
\newcount\MaxItems      
\newcount\StackPointer
\newcount\Point         
\newcount\NextFigure    
\newcount\NextTable     
\newcount\NextItem      

\newcount\StatusStack   
\newcount\NumStack      
\newcount\TypeStack     
\newcount\SpanStack     
\newcount\BoxStack      

\newcount\ItemSTATUS    
\newcount\ItemNUMBER    
\newcount\ItemTYPE      
\newcount\ItemSPAN      
\newbox\ItemBOX         
\newdimen\ItemSIZE      

\newdimen\PageHeight    
\newdimen\TextLeading   
\newdimen\Feathering    
\newcount\LinesPerPage  
\newdimen\ColumnWidth   
\newdimen\ColumnGap     
\newdimen\PageWidth     
\newdimen\BodgeHeight   
\newcount\Leading       

\newdimen\ZoneBSize  
\newdimen\TextSize   
\newbox\ZoneABOX     
\newbox\ZoneBBOX     
\newbox\ZoneCBOX     

\newif\ifFirstSingleItem
\newif\ifFirstZoneA
\newif\ifMakePageInComplete
\newif\ifMoreFigures \MoreFiguresfalse 
\newif\ifMoreTables  \MoreTablesfalse  

\newif\ifFigInZoneB 
\newif\ifFigInZoneC 
\newif\ifTabInZoneB 
\newif\ifTabInZoneC

\newif\ifZoneAFullPage

\newbox\MidBOX    
\newbox\LeftBOX
\newbox\RightBOX
\newbox\PageBOX   

\newif\ifLeftCOL  
\LeftCOLtrue

\newdimen\ZoneBAdjust

\newcount\ItemFits
\def\Yes{1}
\def\No{2}




\MaxItems=15
\NextFigure=0        
\NextTable=1

\BodgeHeight=6pt
\TextLeading=11pt    
\Leading=11
\Feathering=0pt      
\LinesPerPage=61     
\topskip=\TextLeading
\ColumnWidth=20pc    
\ColumnGap=2pc       

\def\ItemSep{\vskip \TextLeading plus \TextLeading minus 4pt}

\FigureBoxesfalse 

\parskip=0pt
\parindent=18pt
\widowpenalty=0
\clubpenalty=10000
\tolerance=1500
\hbadness=1500
\abovedisplayskip=6pt plus 2pt minus 2pt
\belowdisplayskip=6pt plus 2pt minus 2pt
\abovedisplayshortskip=6pt plus 2pt minus 2pt
\belowdisplayshortskip=6pt plus 2pt minus 2pt

\PageHeight=\TextLeading 
\multiply\PageHeight by \LinesPerPage
\advance\PageHeight by \topskip

\PageWidth=2\ColumnWidth
\advance\PageWidth by \ColumnGap




\newcount\DUMMY \StatusStack=\allocationnumber
\newcount\DUMMY \newcount\DUMMY \newcount\DUMMY 
\newcount\DUMMY \newcount\DUMMY \newcount\DUMMY 
\newcount\DUMMY \newcount\DUMMY \newcount\DUMMY
\newcount\DUMMY \newcount\DUMMY \newcount\DUMMY 
\newcount\DUMMY \newcount\DUMMY \newcount\DUMMY

\newcount\DUMMY \NumStack=\allocationnumber
\newcount\DUMMY \newcount\DUMMY \newcount\DUMMY 
\newcount\DUMMY \newcount\DUMMY \newcount\DUMMY 
\newcount\DUMMY \newcount\DUMMY \newcount\DUMMY 
\newcount\DUMMY \newcount\DUMMY \newcount\DUMMY 
\newcount\DUMMY \newcount\DUMMY \newcount\DUMMY

\newcount\DUMMY \TypeStack=\allocationnumber
\newcount\DUMMY \newcount\DUMMY \newcount\DUMMY 
\newcount\DUMMY \newcount\DUMMY \newcount\DUMMY 
\newcount\DUMMY \newcount\DUMMY \newcount\DUMMY 
\newcount\DUMMY \newcount\DUMMY \newcount\DUMMY 
\newcount\DUMMY \newcount\DUMMY \newcount\DUMMY

\newcount\DUMMY \SpanStack=\allocationnumber
\newcount\DUMMY \newcount\DUMMY \newcount\DUMMY 
\newcount\DUMMY \newcount\DUMMY \newcount\DUMMY 
\newcount\DUMMY \newcount\DUMMY \newcount\DUMMY 
\newcount\DUMMY \newcount\DUMMY \newcount\DUMMY 
\newcount\DUMMY \newcount\DUMMY \newcount\DUMMY

\newbox\DUMMY   \BoxStack=\allocationnumber
\newbox\DUMMY   \newbox\DUMMY \newbox\DUMMY 
\newbox\DUMMY   \newbox\DUMMY \newbox\DUMMY 
\newbox\DUMMY   \newbox\DUMMY \newbox\DUMMY 
\newbox\DUMMY   \newbox\DUMMY \newbox\DUMMY 
\newbox\DUMMY   \newbox\DUMMY \newbox\DUMMY

\def\wlog{\immediate\write-1}


\def\GetItemAll#1{%
 \GetItemSTATUS{#1}
 \GetItemNUMBER{#1}
 \GetItemTYPE{#1}
 \GetItemSPAN{#1}
 \GetItemBOX{#1}
}

\def\GetItemSTATUS#1{%
 \Point=\StatusStack
 \advance\Point by #1
 \global\ItemSTATUS=\count\Point
}

\def\GetItemNUMBER#1{%
 \Point=\NumStack
 \advance\Point by #1
 \global\ItemNUMBER=\count\Point
}

\def\GetItemTYPE#1{%
 \Point=\TypeStack
 \advance\Point by #1
 \global\ItemTYPE=\count\Point
}

\def\GetItemSPAN#1{%
 \Point\SpanStack
 \advance\Point by #1
 \global\ItemSPAN=\count\Point
}

\def\GetItemBOX#1{%
 \Point=\BoxStack
 \advance\Point by #1
 \global\setbox\ItemBOX=\vbox{\copy\Point}
 \global\ItemSIZE=\ht\ItemBOX
 \global\advance\ItemSIZE by \dp\ItemBOX
 \TEMPCOUNT=\ItemSIZE
 \divide\TEMPCOUNT by \Leading
 \divide\TEMPCOUNT by 65536
 \advance\TEMPCOUNT by 1
 \ItemSIZE=\TEMPCOUNT pt
 \global\multiply\ItemSIZE by \Leading
}


\def\JoinStack{%
 \ifnum\LengthOfStack=\MaxItems 
  \Warn{WARNING: Stack is full...some items will be lost!}
 \else
  \Point=\StatusStack
  \advance\Point by \LengthOfStack
  \global\count\Point=\ItemSTATUS
  \Point=\NumStack
  \advance\Point by \LengthOfStack
  \global\count\Point=\ItemNUMBER
  \Point=\TypeStack
  \advance\Point by \LengthOfStack
  \global\count\Point=\ItemTYPE
  \Point\SpanStack
  \advance\Point by \LengthOfStack
  \global\count\Point=\ItemSPAN
  \Point=\BoxStack
  \advance\Point by \LengthOfStack
  \global\setbox\Point=\vbox{\copy\ItemBOX}
  \global\advance\LengthOfStack by 1
  \ifnum\ItemTYPE=\Figure 
   \global\MoreFigurestrue
  \else
   \global\MoreTablestrue
  \fi
 \fi
}


\def\LeaveStack#1{%
 {\Iteration=#1
 \loop
 \ifnum\Iteration<\LengthOfStack
  \advance\Iteration by 1
  \GetItemSTATUS{\Iteration}
   \advance\Point by -1
   \global\count\Point=\ItemSTATUS
  \GetItemNUMBER{\Iteration}
   \advance\Point by -1
   \global\count\Point=\ItemNUMBER
  \GetItemTYPE{\Iteration}
   \advance\Point by -1
   \global\count\Point=\ItemTYPE
  \GetItemSPAN{\Iteration}
   \advance\Point by -1
   \global\count\Point=\ItemSPAN
  \GetItemBOX{\Iteration}
   \advance\Point by -1
   \global\setbox\Point=\vbox{\copy\ItemBOX}
 \repeat}
 \global\advance\LengthOfStack by -1
}


\newif\ifStackNotClean

\def\CleanStack{%
 \StackNotCleantrue
 {\Iteration=0
  \loop
   \ifStackNotClean
    \GetItemSTATUS{\Iteration}
    \ifnum\ItemSTATUS=\InStack
     \advance\Iteration by 1
     \else
      \LeaveStack{\Iteration}
    \fi
   \ifnum\LengthOfStack<\Iteration
    \StackNotCleanfalse
   \fi
 \repeat}
}


\def\FindItem#1#2{%
 \global\StackPointer=-1 
 {\Iteration=0
  \loop
  \ifnum\Iteration<\LengthOfStack
   \GetItemSTATUS{\Iteration}
   \ifnum\ItemSTATUS=\InStack
    \GetItemTYPE{\Iteration}
    \ifnum\ItemTYPE=#1
     \GetItemNUMBER{\Iteration}
     \ifnum\ItemNUMBER=#2
      \global\StackPointer=\Iteration
      \Iteration=\LengthOfStack 
     \fi
    \fi
   \fi
  \advance\Iteration by 1
 \repeat}
}


\def\FindNext{%
 \global\StackPointer=-1 
 {\Iteration=0
  \loop
  \ifnum\Iteration<\LengthOfStack
   \GetItemSTATUS{\Iteration}
   \ifnum\ItemSTATUS=\InStack
    \GetItemTYPE{\Iteration}
   \ifnum\ItemTYPE=\Figure
    \ifMoreFigures
      \global\NextItem=\Figure
      \global\StackPointer=\Iteration
      \Iteration=\LengthOfStack 
    \fi
   \fi
   \ifnum\ItemTYPE=\Table
    \ifMoreTables
      \global\NextItem=\Table
      \global\StackPointer=\Iteration
      \Iteration=\LengthOfStack 
    \fi
   \fi
  \fi
  \advance\Iteration by 1
 \repeat}
}


\def\ChangeStatus#1#2{%
 \Point=\StatusStack
 \advance\Point by #1
 \global\count\Point=#2
}



\def\Zone{\InZoneA}

\ZoneBAdjust=0pt

\def\MakePage{
 \global\ZoneBSize=\PageHeight
 \global\TextSize=\ZoneBSize
 \global\ZoneAFullPagefalse
 \global\topskip=\TextLeading
 \MakePageInCompletetrue
 \MoreFigurestrue
 \MoreTablestrue
 \FigInZoneBfalse
 \FigInZoneCfalse
 \TabInZoneBfalse
 \TabInZoneCfalse
 \global\FirstSingleItemtrue
 \global\FirstZoneAtrue
 \global\setbox\ZoneABOX=\box\VOIDBOX
 \global\setbox\ZoneBBOX=\box\VOIDBOX
 \global\setbox\ZoneCBOX=\box\VOIDBOX
 \loop
  \ifMakePageInComplete
 \FindNext
 \ifnum\StackPointer=-1
  \NextItem=-1
  \MoreFiguresfalse
  \MoreTablesfalse
 \fi
 \ifnum\NextItem=\Figure
   \FindItem{\Figure}{\NextFigure}
   \ifnum\StackPointer=-1 \global\MoreFiguresfalse
   \else
    \GetItemSPAN{\StackPointer}
    \ifnum\ItemSPAN=\Single \def\Zone{\InZoneB}\relax
     \ifFigInZoneC \global\MoreFiguresfalse\fi
    \else
     \def\Zone{\InZoneA}
     \ifFigInZoneB \def\Zone{\InZoneC}\fi
    \fi
   \fi
   \ifMoreFigures\Print{}\FigureItems\fi
 \fi
\ifnum\NextItem=\Table
   \FindItem{\Table}{\NextTable}
   \ifnum\StackPointer=-1 \global\MoreTablesfalse
   \else
    \GetItemSPAN{\StackPointer}
    \ifnum\ItemSPAN=\Single\relax
     \ifTabInZoneC \global\MoreTablesfalse\fi
    \else
     \def\Zone{\InZoneA}
     \ifTabInZoneB \def\Zone{\InZoneC}\fi
    \fi
   \fi
   \ifMoreTables\Print{}\TableItems\fi
 \fi
   \MakePageInCompletefalse 
   \ifMoreFigures\MakePageInCompletetrue\fi
   \ifMoreTables\MakePageInCompletetrue\fi
 \repeat
 \ifZoneAFullPage
  \global\TextSize=0pt
  \global\ZoneBSize=0pt
  \global\vsize=0pt\relax
  \global\topskip=0pt\relax
  \vbox to 0pt{\vss}
  \eject
 \else
 \global\advance\ZoneBSize by -\ZoneBAdjust
 \global\vsize=\ZoneBSize
 \global\hsize=\ColumnWidth
 \global\ZoneBAdjust=0pt
 \ifdim\TextSize<23pt
 \Warn{}
 \Warn{* Making column fall short: TextSize=\the\TextSize *}
 \vskip-\lastskip\eject\fi
 \fi
}

\def\MakeRightCol{
 \global\TextSize=\ZoneBSize
 \MakePageInCompletetrue
 \MoreFigurestrue
 \MoreTablestrue
 \global\FirstSingleItemtrue
 \global\setbox\ZoneBBOX=\box\VOIDBOX
 \def\Zone{\InZoneB}
 \loop
  \ifMakePageInComplete
 \FindNext
 \ifnum\StackPointer=-1
  \NextItem=-1
  \MoreFiguresfalse
  \MoreTablesfalse
 \fi
 \ifnum\NextItem=\Figure
   \FindItem{\Figure}{\NextFigure}
   \ifnum\StackPointer=-1 \MoreFiguresfalse
   \else
    \GetItemSPAN{\StackPointer}
    \ifnum\ItemSPAN=\Double\relax
     \MoreFiguresfalse\fi
   \fi
   \ifMoreFigures\Print{}\FigureItems\fi
 \fi
 \ifnum\NextItem=\Table
   \FindItem{\Table}{\NextTable}
   \ifnum\StackPointer=-1 \MoreTablesfalse
   \else
    \GetItemSPAN{\StackPointer}
    \ifnum\ItemSPAN=\Double\relax
     \MoreTablesfalse\fi
   \fi
   \ifMoreTables\Print{}\TableItems\fi
 \fi
   \MakePageInCompletefalse 
   \ifMoreFigures\MakePageInCompletetrue\fi
   \ifMoreTables\MakePageInCompletetrue\fi
 \repeat
 \ifZoneAFullPage
  \global\TextSize=0pt
  \global\ZoneBSize=0pt
  \global\vsize=0pt\relax
  \global\topskip=0pt\relax
  \vbox to 0pt{\vss}
  \eject
 \else
 \global\vsize=\ZoneBSize
 \global\hsize=\ColumnWidth
 \ifdim\TextSize<23pt
 \Warn{}
 \Warn{* Making column fall short: TextSize=\the\TextSize *}
 \vskip-\lastskip\eject\fi
\fi
}

\def\FigureItems{
 \Print{Considering...}
 \ShowItem{\StackPointer}
 \GetItemBOX{\StackPointer} 
 \GetItemSPAN{\StackPointer}
  \CheckFitInZone 
  \ifnum\ItemFits=\Yes
   \ifnum\ItemSPAN=\Single
     \ChangeStatus{\StackPointer}{\InZoneB} 
     \global\FigInZoneBtrue
     \ifFirstSingleItem
      \hbox{}\vskip-\BodgeHeight
     \global\advance\ItemSIZE by \TextLeading
     \fi
     \unvbox\ItemBOX\ItemSep
     \global\FirstSingleItemfalse
     \global\advance\TextSize by -\ItemSIZE
     \global\advance\TextSize by -\TextLeading
   \else
    \ifFirstZoneA
     \global\advance\ItemSIZE by \TextLeading
     \global\FirstZoneAfalse\fi
    \global\advance\TextSize by -\ItemSIZE
    \global\advance\TextSize by -\TextLeading
    \global\advance\ZoneBSize by -\ItemSIZE
    \global\advance\ZoneBSize by -\TextLeading
    \ifFigInZoneB\relax
     \else
     \ifdim\TextSize<3\TextLeading
     \global\ZoneAFullPagetrue
     \fi
    \fi
    \ChangeStatus{\StackPointer}{\Zone}
    \ifnum\Zone=\InZoneC \global\FigInZoneCtrue\fi
  \fi
   \Print{TextSize=\the\TextSize}
   \Print{ZoneBSize=\the\ZoneBSize}
  \global\advance\NextFigure by 1
   \Print{This figure has been placed.}
  \else
   \Print{No space available for this figure...holding over.}
   \Print{}
   \global\MoreFiguresfalse
  \fi
}

\def\TableItems{
 \Print{Considering...}
 \ShowItem{\StackPointer}
 \GetItemBOX{\StackPointer} 
 \GetItemSPAN{\StackPointer}
  \CheckFitInZone 
  \ifnum\ItemFits=\Yes
   \ifnum\ItemSPAN=\Single
    \ChangeStatus{\StackPointer}{\InZoneB}
     \global\TabInZoneBtrue
     \ifFirstSingleItem
      \hbox{}\vskip-\BodgeHeight
     \global\advance\ItemSIZE by \TextLeading
     \fi
     \unvbox\ItemBOX\ItemSep
     \global\FirstSingleItemfalse
     \global\advance\TextSize by -\ItemSIZE
     \global\advance\TextSize by -\TextLeading
   \else
    \ifFirstZoneA
    \global\advance\ItemSIZE by \TextLeading
    \global\FirstZoneAfalse\fi
    \global\advance\TextSize by -\ItemSIZE
    \global\advance\TextSize by -\TextLeading
    \global\advance\ZoneBSize by -\ItemSIZE
    \global\advance\ZoneBSize by -\TextLeading
    \ifFigInZoneB\relax
     \else
     \ifdim\TextSize<3\TextLeading
     \global\ZoneAFullPagetrue
     \fi
    \fi
    \ChangeStatus{\StackPointer}{\Zone}
    \ifnum\Zone=\InZoneC \global\TabInZoneCtrue\fi
   \fi
  \global\advance\NextTable by 1
   \Print{This table has been placed.}
  \else
  \Print{No space available for this table...holding over.}
   \Print{}
   \global\MoreTablesfalse
  \fi
}


\def\CheckFitInZone{%
{\advance\TextSize by -\ItemSIZE
 \advance\TextSize by -\TextLeading
 \ifFirstSingleItem
  \advance\TextSize by \TextLeading
 \fi
 \ifnum\Zone=\InZoneA\relax
  \else \advance\TextSize by -\ZoneBAdjust
 \fi
 \ifdim\TextSize<3\TextLeading \global\ItemFits=\No
 \else \global\ItemFits=\Yes\fi}
}

\def\BF#1#2{
 \ItemSTATUS=\InStack
 \ItemNUMBER=#1
 \ItemTYPE=\Figure
 \if#2S \ItemSPAN=\Single
  \else \ItemSPAN=\Double
 \fi
 \setbox\ItemBOX=\vbox{}
}

\def\BT#1#2{
 \ItemSTATUS=\InStack
 \ItemNUMBER=#1
 \ItemTYPE=\Table
 \if#2S \ItemSPAN=\Single
  \else \ItemSPAN=\Double
 \fi
 \setbox\ItemBOX=\vbox{}
}

\def\BeginOpening{%
 \hsize=\PageWidth
 \global\setbox\ItemBOX=\vbox\bgroup
}

\let\begintopmatter=\BeginOpening  

\def\EndOpening{%
 \egroup
 \ItemNUMBER=0
 \ItemTYPE=\Figure
 \ItemSPAN=\Double
 \ItemSTATUS=\InStack
 \JoinStack
}

%

\newbox\tmpbox

\def\FC#1#2#3#4{%
  \ItemSTATUS=\InStack
  \ItemNUMBER=#1
  \ItemTYPE=\Figure
  \if#2S
    \ItemSPAN=\Single \TEMPDIMEN=\ColumnWidth
  \else
    \ItemSPAN=\Double \TEMPDIMEN=\PageWidth
  \fi
  {\hsize=\TEMPDIMEN
   \global\setbox\ItemBOX=\vbox{%
     \ifFigureBoxes
       \B{\TEMPDIMEN}{#3}
     \else
       \vbox to #3{\vfil}%
     \fi%
     \eightpoint\rm\bls{\rTenPT}%
     \vskip 5.5pt plus 6pt%
     \setbox\tmpbox=\vbox{#4\par}%
     \ifdim\ht\tmpbox>10pt 
       \noindent #4\par%
     \else
       \hbox to \hsize{\hfil #4\hfil}%
     \fi%
   }%
  }%
  \JoinStack%
  \Print{Processing source for figure {\the\ItemNUMBER}}%
}

\let\figure=\FC  

\def\TH#1#2#3#4{%
 \ItemSTATUS=\InStack
 \ItemNUMBER=#1
 \ItemTYPE=\Table
 \if#2S \ItemSPAN=\Single \TEMPDIMEN=\ColumnWidth
  \else \ItemSPAN=\Double \TEMPDIMEN=\PageWidth
 \fi
{\hsize=\TEMPDIMEN
\eightpoint\bls{\rTenPT}\rm
\global\setbox\ItemBOX=\vbox{\noindent#3\vskip 5.5pt plus5.5pt\noindent#4}}
 \JoinStack
 \Print{Processing source for table {\the\ItemNUMBER}}
}

\let\table=\TH  

\def\UnloadZoneA{%
\FirstZoneAtrue
 \Iteration=0
  \loop
   \ifnum\Iteration<\LengthOfStack
    \GetItemSTATUS{\Iteration}
    \ifnum\ItemSTATUS=\InZoneA
     \GetItemBOX{\Iteration}
     \ifFirstZoneA \vbox to \BodgeHeight{\vfil}%
     \FirstZoneAfalse\fi
     \unvbox\ItemBOX\ItemSep
     \LeaveStack{\Iteration}
     \else
     \advance\Iteration by 1
   \fi
 \repeat
}

\def\UnloadZoneC{%
\Iteration=0
  \loop
   \ifnum\Iteration<\LengthOfStack
    \GetItemSTATUS{\Iteration}
    \ifnum\ItemSTATUS=\InZoneC
     \GetItemBOX{\Iteration}
     \ItemSep\unvbox\ItemBOX
     \LeaveStack{\Iteration}
     \else
     \advance\Iteration by 1
   \fi
 \repeat
}


\def\ShowItem#1{
  {\GetItemAll{#1}
  \Print{\the#1:
  {TYPE=\ifnum\ItemTYPE=\Figure Figure\else Table\fi}
  {NUMBER=\the\ItemNUMBER}
  {SPAN=\ifnum\ItemSPAN=\Single Single\else Double\fi}
  {SIZE=\the\ItemSIZE}}}
}

\def\ShowStack{%
 \Print{}
 \Print{LengthOfStack = \the\LengthOfStack}
 \ifnum\LengthOfStack=0 \Print{Stack is empty}\fi
 \Iteration=0
 \loop
 \ifnum\Iteration<\LengthOfStack
  \ShowItem{\Iteration}
  \advance\Iteration by 1
 \repeat
}

\def\B#1#2{%
\hbox{\vrule\kern-0.4pt\vbox to #2{%
\hrule width #1\vfill\hrule}\kern-0.4pt\vrule}
}

\def\Ref#1{\begingroup\global\setbox\TEMPBOX=\vbox{\hsize=2in\noindent#1}\endgroup
\ht1=0pt\dp1=0pt\wd1=0pt\vadjust{\vtop to 0pt{\advance
\hsize0.5pc\kern-10pt\moveright\hsize\box\TEMPBOX\vss}}}

\def\MarkRef#1{\leavevmode\thinspace\hbox{\vrule\vtop
{\vbox{\hrule\kern1pt\hbox{\vphantom{\rm/}\thinspace{\rm#1}%
\thinspace}}\kern1pt\hrule}\vrule}\thinspace}%


\output{%
 \ifLeftCOL
  \global\setbox\LeftBOX=\vbox to \ZoneBSize{\box255\unvbox\ZoneBBOX}
  \global\LeftCOLfalse
  \MakeRightCol
 \else
  \setbox\RightBOX=\vbox to \ZoneBSize{\box255\unvbox\ZoneBBOX}
  \setbox\MidBOX=\hbox{\box\LeftBOX\hskip\ColumnGap\box\RightBOX}
  \setbox\PageBOX=\vbox to \PageHeight{%
  \UnloadZoneA\box\MidBOX\UnloadZoneC}
  \shipout\vbox{\PageHead\box\PageBOX\PageFoot}
  \global\advance\pageno by 1
  \global\HeaderNumber=\DefaultHeader
  \global\LeftCOLtrue
  \CleanStack
  \MakePage
 \fi
}


\catcode `\@=12 


\def\solar{\ifmmode_{\mathord\odot}\else$_{\mathord\odot}$\fi}     
\def\arcs{\ifmmode {'' }\else $'' $\fi}     
\def\arcm{\ifmmode {' }\else $' $\fi}     
\def\arcmper{\ifmmode\rlap.{' }\else $\rlap{.}' $\fi} 
\def\arcsper{\ifmmode \rlap.{'' }\else $\rlap{.}'' $\fi} 
\def\et{et\thinspace al.\ }    

\Autonumber 
\begintopmatter 
\title{NGC 6397: A case study in the resolution of post-collapse
globular cluster cores}
\author{G. A. Drukier}  
\affiliation{Institute of Astronomy,  Madingley Rd., Cambridge, CB3 0HA} 
\affiliation{Internet: drukier@mail.ast.cam.ac.uk}
\shortauthor{G. A. Drukier} 
\shorttitle{Resolution of globular cluster cores}

\abstract 
Model surface brightness profiles based on Fokker-Planck simulations
have been used to assess the high resolution surface brightness profile
of the globular cluster  NGC 6397 of Lauzeral \et (1992).  The profile
is well fitted by the model in the maximum-expansion phase of a
gravothermal oscillation with a core radius of 0.06 pc (6\arcs), but an
unresolved core cannot be ruled out. A core this size is also expected
in highly evolved clusters including significant numbers of primordial
binaries.  A more massive cluster will provide better statistics, as
would star counts.  The greater distance of the other globular clusters
and the requirements of star counting will demand very high  resolution
observations to definitively resolve cores this size.

\keywords{globular clusters: individual: NGC 6397---globular clusters: general} 

\maketitle

\section{Introduction} 
\tx 
In globular clusters which have passed through core collapse, the
central density profile takes the form of a power-law.  At some point
the profile must flatten out, forming a core of approximately constant
density, but whether this flattening is observable will depend on the
underlying stellar distribution function and the sampling of this
distribution function by the finite number of stars.  From a
theoretical perspective, the size of the post-collapse core of a
cluster will depend on the nature of the heating mechanism which drives
the re-expansion.  Models which depend on binaries resulting from
three-body encounters (Lee 1987) require much higher central densities
before contraction is  reversed than do models using tidally-formed
binaries (Statler, Ostriker \& Cohn, 1987) or with primordial binaries
(Gao \et 1991). Further, if gravothermal oscillations, which are
observed in statistical numerical models such as gas sphere or
Fokker-Planck codes, also occur in nature, then the post-collapse core
would be much larger than is expected from a steady expansion. In the
case of gravothermal oscillations the core radius is of order 1\% of
the half-mass radius of the cluster at maximum expansion (Murphy, Cohn
\& Hut 1990); from a few hundredths to  a tenth of a parsec. A core
this large may be observable under some conditions and its detection or
non-detection would help in guiding further theoretical work. The
finite number of stars in the region of interest places a fundamental
restriction on the precision with which the comparisons can be made.

Lauer \et (1991) analyzed a $U$-band HST image of the
high-concentration cluster M15. After removing the bright stars, they
claimed to have resolved a 2\arcsper 2 core in the  residual light
profile. Grabhorn \et (1992) fitted a Fokker-Planck model to the HST
profile and stated that the observations were best matched by the model
when it was in the state of maximum expansion. However, further
consideration of the relationship between the models and the
observations by Grabhorn \et (1993) has demonstrated that for these M15
data it is in practice impossible to tell the difference between the
two extreme phases of gravothermal oscillations.  This is because there
are so few stars in the region of interest that sampling variations
mask the difference.  Monte Carlo representations of the light profile
drawn from the distribution functions of the two model phases overlap
to the extent that stellar  distributions drawn from the
maximum-expansion distribution can look cuspy, and vice versa.

Yanny \et (1993) have done photometry on new HST $V$ and $I$
observations of the core of M15.  They find that the star counts are
consistent with both a power-law cusp and with a small core of radius
less than 1\arcsper 5 (0.09 pc). They showed that given the large wings
of the HST point spread function, the residual light profile is an
unreliable guide to the  cluster's structure. A nearer cluster is
required to resolve the question of the observability of such a
cluster's core.

The recent high resolution observation by Lauzeral \et (1992; hereafter
LOAM) of the surface brightness profile of the nearby cluster NGC 6397
goes some way towards clarifying this issue.  NGC 6397 is in many ways
the key cluster in this discussion. First, it has a central cusp in its
surface brightness and surface density. This is generally taken as
signifying that a cluster has passed through core collapse. Second, at
the distance of NGC 6397 (2.2 kpc; Drukier \et 1993), an arc second
corresponds to 0.01 pc, and a maximum-expansion-phase core should be
easily resolvable.  Third, there is an extensive set  of other
observations available to help constrain models of the cluster (Fahlman
\et 1989, Drukier \et 1993).

One of the problems which have arisen in understanding the cores of
cusp-profile clusters is the origin of the colour gradients observed in
them.  Djorgovski \et (1991) have argued that the colour gradients are
due to a deficiency of red giants in the cluster cores. It is this
hypothesis that LOAM have investigated with their new observations.
These are based on data taken with the Danish 1.5 m telescope at ESO
under conditions of 0\arcsper 95 seeing.  What LOAM demonstrate is that
the colour gradient in NGC 6397 can be eliminated if the light from the
brighter red giants and from  the blue stragglers is removed.  The
resulting surface brightness profile then loses its power-law shape in
the central few arc seconds and flattens out.  In the final profile
LOAM find the core radius to be 6\arcs. With the removal of the
brightest stars the surface brightness is dominated by the less evolved
stars and so should better reflect the underlying density
distribution.

The comparison of these data with a Fokker-Planck model which includes
gravothermal oscillations, will both investigate whether NGC 6397 is in
a maximum expansion phase (as is expected) and address the question of
the limits placed on the observability of post-collapse cores by the
finite number of stars. A distinction should be made between a surface
brightness profile, which is derived from the integrated light of the
individual stars, and a surface density profile, which comes from star
counts.  When a surface brightness profile is to be compared with a
model, the mass--luminosity relation is needed to convert between the
observed surface brightness profile and the model profile which is
expressed in terms of the number of stars per unit area.

\section{ Model comparison}   
\tx  

As part of a project to compare Fokker-Planck models with observations
of globular clusters (Drukier 1993), I discovered one model in
particular for NGC 6397 which gave a good match to  mass functions at
three radii as well as the surface density profile of the cluster's
bright stars.  Due to the crowding of stellar images in the core of the
cluster a relatively bright magnitude cutoff had to be used (Drukier
\et 1993), hence the central region of the observed  surface density
profile has too few stars to resolve anything within the central
10\arcs. The LOAM surface brightness profile allows the extension of
this comparison into the very centre of the cluster.

The models  are based on the numerical integration of the isotropic,
orbit-averaged, Fokker-Planck equation, as originally described in Cohn
(1980). This is a statistical approach with the physical state of the
cluster being given by a distribution function in energy space.  The
stellar mass spectrum is represented by a series of  mass bins, each
with its own distribution function.

A tidal boundary has been applied in energy space following the method
of Lee \& Ostriker (1987).  Given an initial tidal radius, the mean
density of the cluster within that radius is easily determined. At
subsequent times, the tidal boundary in energy is given by the
potential at the radius enclosing the same mean density. A fraction of
the distribution function at energies greater than this ``tidal
energy'' is removed at each time step. The fraction removed at a given
energy is determined by the tidal stripping timescale and by the energy
difference with respect to the tidal energy.  The stripping rate is
proportional to the cube of this energy difference.

In order to drive the gravothermal oscillations, an energy source is
needed to oppose the gravitational collapse. In a star the energy comes
from nuclear fusion. In a globular cluster it is expected that the
energy comes from the hardening of binary stars.  As interactions with
the rest of the stars in the cluster cause the binary to become more
tightly bound, energy is liberated to reverse the collapse.  In the
models used here, the binaries are assumed to form via three-body
interactions. This energy source is represented statistically by
calculating the binary creation rate and the average amount of energy
released by each binary.

These models are much the same as those in Drukier, Fahlman, \& Richer
(1992), but  two additions have been made. First, the effects of
stellar  mass loss (due to stellar evolution) have been included by
allowing the masses of the various mass bins to vary with time in much
the same manner as in Chernoff \& Weinberg (1990). This leads to a much
more realistic handling of the mass spectrum and remnant population
than when no stellar evolution is included (as, for  example, in
Drukier \et 1992). It also restricts the choice of initial mass
function to ones which do not result in so much mass loss that the
cluster would not survive to the present. This class of model is
discussed further in Drukier (1993).  At any given time, there is one
mass bin with  a changing mass. This bin, which I will refer to as the
``evolving bin'', represents the stars which currently are near to the
turn-off or have evolved past it onto the giant and horizontal branches
or their final states as white dwarfs or neutron stars.  A non-evolving
bin represents stars which are still somewhere on the main sequence or
which have finished evolving  and are now degenerate remnants.

The other difference is that for the model described here the time step
has been taken to be two central relaxation times.  Such a small time
step allows gravothermal oscillations to take place. By way of
comparison, a model with large time steps (typically several thousand
central relaxation times in the post-collapse phase) was also run. The
evolution of the core radius for both models is shown in  Fig.~1.  The
curve for the non-oscillating model has been shifted 0.16 Gyr earlier
to match the time of core collapse. Since the details of the evolution
are dependent on the initial conditions and the choice of time step,
these models should not be taken as claiming a dynamical age for NGC
6397.

\figure{1}{S}{85mm}{ 
\bf Figure 1. \rm  Core radius as a function of time for the model
described here.  The inset shows the core radius for the whole run,
while the main part of the figure shows the post--core-collapse epoch
to emphasize the gravothermal oscillations. The dashed line corresponds
to the evolution of the same model when large time steps, which
suppress gravothermal oscillations, are used. The curve for the model
without oscillations has been shifted 0.16 Gyr earlier in order to
match the core collapse times. The two vertical bars indicate the times
of the two profiles shown in Fig.~2. }

The initial model contained  25  bins covering the initial mass range
between 0.1 and 20$M\solar$.  The initial numbers of stars in each bin
were distributed as two power-laws, one, for stars with $M<0.4M\solar$,
with mass spectral index $x=1.5$ (where $x=1.35$ for a Salpeter mass
function), the other, for $M>0.4M\solar$, had $x=0.9$ . The mass
function was made continuous at $0.4M\solar$. The stellar evolution
times and final masses  were the same as those used by Chernoff \&
Weinberg (1990) except that stellar evolution was assumed to stop at 12
Gyr. This was done to keep the mean mass of the stars in the  bin which
was to start evolving at that time constant for the purposes of
comparison. By doing this, the question of the appropriate
light-to-mass ratio to use for comparison to the surface brightness
profile need only be addressed once and not for each time of
comparison. One limitation of the approach used here to handle stellar
evolution is that while a bin is evolving it is held to contain a
mixture of main-sequence stars and  degenerate remnants. The cutoff at
12 Gyr, being the assumed lifetime of a star with the mass of the bin
boundary, removes this inconsistency from the comparison.  On the other
hand, this limit results in another inconsistency, one between the mass
of this bin, which is the one contributing most of the light, and that
of the stars presently observed at the turnoff and above in NGC 6397.
The small mass difference should not have a large impact on the
dynamics.  At such late times the effects of stellar evolution are
quite small, so the lack of the heating due to stellar evolution should
also not be a concern. The initial distribution function was that
appropriate to a King (1966) model with $W_0=6$ with total mass
$7.5\times 10^5 M\solar$ and limiting radius 39.1 pc.  The initial
tidal radius was also set to be 39.1 pc.

The model was run for $10^5$ time steps taking about 12 cpu-days on  a
SparcStation IPC.  Six major re-expansions were  observed in the 600
Myr span covered by the post-core-collapse  phase shown in Fig.~1.   In
order to contrast the possible extremes in the size of a post-collapse
core, I will be using the two extreme extrema indicated in Fig.~1 for
the comparisons.

For comparison with the observations the model densities were first
projected and then integrated over annuli corresponding to the  LOAM
data points.  The projected number densities at each radius were then
converted to surface brightnesses by adopting appropriate magnitudes
for each mass bin making a significant contribution to the cluster
luminosity.  There are several concerns which must be addressed in
choosing these magnitudes. The first is the importance of the
contribution of the low mass stars to the overall light of the
cluster.  For these low mass stars, $\log L \sim 2.5 \log M + {\rm
constant}$ (Bergbusch \& VandenBerg 1992). If the local mass function
has $x<1.5$, then the cumulative contribution of these stars converges.
If $x$  is much smaller then the total light from the faint stars is
unimportant.  In the model presented here, mass segregation, the
enhancement of higher mass stars with respect to lower mass stars in
the centres of clusters due to equipartition of energy, reduces the
mass spectral index of the low mass stars from its initial value of
$x=1.5$ to $x=0.8$ in the region we are interested in. Therefore we
should be able to safely neglect the lowest mass bins.

How many bins are actually needed was determined in the following
manner.  LOAM removed the blue stragglers and the stars with magnitudes
brighter than $V=14.0$. I estimated the mean magnitude for the stars
between  $V=14.0$ and $V=16.4$ using the luminosity function of Alcaino
\et (1987). This came  out to be 15.5, but  it is unclear whether the
$V=16.4$ magnitude cut off is appropriate for the turn-off bin. The
reason for this is the mismatch mentioned above between the mass of
this bin and that of the turn-off stars in the cluster.  Further, there
is a mismatch in age and metallicity between the isochrone used to
assign the  evolutionary time scale of the mass bins in the cluster
model and the observed age and metallicity of NGC 6397. In any event,
first estimates for the the mean magnitudes of the next three mass
bins  were take from the  isochrone which is from VandenBerg and Bell
(1985).

The gravothermal oscillations in this model affect only the inner 0.1
pc, so I attempted to improve these magnitude estimates by fitting the
two model profiles to the  eight outermost LOAM data points.  These
fits did not provide a strong constraint on the magnitudes. In the case
of the least massive of the four bins a limit of $V>18.$ was found.
This suggests that there are an insufficient number of stars in this
bin to contribute significant light to the surface brightness profile.
This mass bin has an upper mass limit of 0.58 $M\solar$.  Given the
initial mass function for the model and the $Z=10^{-4}$ isochrone from
Bergbusch \& VandenBerg (1992), the total light contributed  from stars
less massive than 0.58 $M\solar$ is only 19\% of the light of all the
stars up to $V=14.0$. Mass segregation reduces this fraction further,
to less than 12\%, so only the turn-off bin and the next two were used
in the comparisons. Based on these fits  mean magnitudes were adopted
for the the three remaining bins, with $V=16.$ being used for the bin
covering the upper main sequence and more evolved stars.   The
estimated uncertainty on these magnitudes is 0.3.

\figure{2}{S}{85mm}{  
\bf Figure 2. 
\rm The data of Lauzeral \et (1992) are compared with the two extreme
model profiles. The solid line shows the core collapse phase and the
dashed line is the maximum expansion phase. The bounded, closely-hashed
regions indicate the $1\sigma$ uncertainties associated with sampling.
The wider hashing, extending beyond the bounding lines,
include the magnitude uncertainties as well. }

A surface brightness profile from a model is based on a continuous
distribution. Assuming that the observed profile has the same
underlying distribution, the observations represent a discrete sampling
of that distribution.  A proper comparison  between the observations
and the models should take into account the sampling uncertainties in
the model ``data points''. Figure~2 presents the comparison between the
observed surface brightness profile and those of the model at the two
extrema.  The points are the LOAM data and the two lines central to the
hashed regions are the model profiles at the two extrema.  The width of
the hashed regions indicates the uncertainties. The closer-spaced
hashing within the bounding lines represent $1\sigma$ Poisson errors on
the integrated model points, while the wider shadings, extending beyond
the bounding lines, include the 0.3 magnitude uncertainties in adopted
magnitudes added in quadrature.

\table{1}{S}{\bf Table 1. \rm Probabilities of getting the observed values of
$\chi^2$.} 
{\tabskip=1em plus2em minus.5em 
\halign to\hsize{#\hfil&#\hfil&#\hfil&#\hfil&#\hfil&#\hfil\cr 
Points$^a$ &  $SB_{bkgd}^b$   & $P(\chi_{cc}^2)^c$  & $P(\chi_{exp}^2)^c$ 
&$P(\chi_{cc}^2)^d$  & $P(\chi_{exp}^2)^d$\cr 
\noalign{\One} 
1--20 & 16.89 &0.18 & 0.96 & 0.81 & 1.00\cr 
1--20 & 18    & 0.19 & 0.97 & 0.80 & 1.00\cr
1--20 & 20    & 0.17 & 0.97 & 0.81 & 1.00\cr 
\noalign{\Half} 
2--20 & 16.89 &0.21 & 0.95 & 0.84 & 0.999\cr 
2--20 & 18    & 0.24 & 0.95 & 0.86 & 0.9995\cr
2--20 & 20    & 0.22 & 0.95 & 0.86 & 1.00\cr 
\noalign{\Half} 
1--12 & 16.89 & 0.21 & 0.97 & 0.57 & 0.998\cr 
1--12 & 18    & 0.21 & 0.97 & 0.57 & 0.995\cr
1--12 & 20    & 0.23 & 0.96 & 0.57 & 0.996\cr 
\noalign{\Half} 
2--12 & 16.89 &0.25 & 0.97 & 0.61 & 0.996\cr 
2--12 & 18    & 0.25 & 0.96 & 0.62 & 0.997\cr
2--12 & 20    & 0.23 & 0.96 & 0.62 & 0.993\cr 
\noalign{\One}
\noalign{$^a$Counting from centre}
\noalign{$^b$Central background surface brightness, see text} 
\noalign{$^c$Sampling error only}
\noalign{$^d$Sampling and magnitude errors}}}

To the eye, the maximum expansion profile provides a much better fit to
the observations than does the core collapse profile.  In order to
quantify this comparison,  I used Monte Carlo techniques to produce
model surface brightness profiles consistent with Poisson statistics in
the densities.  For each mass species contributing to the surface
brightness profile, and for each observing annulus, the number of stars
of that mass ``observed'' in that bin was randomly drawn. These numbers
were then used to calculate the surface brightness in that bin.  This
surface brightness profile was then compared with the parent profile
and a $\chi^2$ value  was calculated using the same uncertainties as
for the observations.  The fraction of representations which had
$\chi^2$  greater than or equal to that observed was then used to give
the quality of the fit.

The first column of Table~1 gives the range of LOAM data points
(counting from the centre) used in each comparison.  The full data set
contains 20 points, of which the outermost eight were used in
estimating the magnitudes.  The reason for starting some comparisons at
the second data point is coupled with the reason for the numbers listed
in column 2. While, as was argued above, the fainter stars, which have
been neglected in producing these profiles, do not contribute a
significant amount of light {\em on average}, the Monte Carlo sampling
can occasionally introduce anomalies. For the innermost point of
maximum-expansion profile there less than three stars in each of the
mass classes. Quite frequently, the simulated counts in this bin would
be zero, giving an undefined surface brightness.  To alleviate this
problem, and to test the assumption that the faint stars are relatively
unimportant, a background surface brightness for the centre was assumed
in these cases and these are listed in column 2. The value of 16.89 is
the observed surface brightness in the bin.  The choice for a
background surface brightness did not have any effect on the
probabilities. Hence the neglect of the faint stars is justified.
Alternatively, we could just ignore the central bin and start at the
second, since the innermost data point only reinforces the trend
already established by those further out. What this demonstrates is
that the poor fit of the core-collapse profile is not exclusively due
to the central datum.  There is little effect except to make the
core-collapse profiles somewhat more probable than the case where the
central point is included; an unsurprising result.

The rightmost two columns of Table~1 repeat these comparisons, but
include the uncertainty in the stellar magnitudes.  Since the
comparison is being made in terms of surface brightness, and since I
have adopted the magnitudes based on fits to the outer part of the
observed surface brightness profile, any systematic uncertainties in
either the distance to NGC 6397 or the magnitudes will balance out.  On
the other hand, the fits do allow for some tradeoff in magnitude
between the three mass bins. This set of comparisons allows for these
uncertainties by assuming that the probability that a given magnitude
is the correct magnitude for each bin is normally distributed about my
adopted magnitudes with a $1\sigma$ dispersion of 0.3 magnitudes.
With this assumption, both model profiles are more probably fit, but
the maximum-expansion phase is still the more strongly favoured. The
reason for the differences in the probabilities is that outermost
points contribute much less strongly to the $\chi^2$ distribution when
the magnitude errors are included.  The high probabilities assigned to
the maximal-expansion model when the magnitude uncertainty is included
suggests that 0.3 magnitudes are a conservative estimate of these
errors.  The two sets of comparisons provide limiting cases on the
magnitude uncertainties.

On the basis of Table~1, the core-collapsed profile cannot be ruled
out, but the maximum expansion phase, with a core radius of  6\arcs
(0.06 pc), is to be preferred.

\section{ Discussion} 
\tx 
Given this probable resolution of a 6\arcs core in NGC 6397, what can
be said about its dynamical state? One possibility is that NGC 6397 had
a population of primordial binaries which have given the cluster a
large core (Gao \et 1992). Alternatively, if NGC 6397 is undergoing
gravothermal oscillations, then it is not so  surprising that we have
caught it in the maximum expansion phase, as can  be seen in Fig.~1.

However,  due to the sampling problems involved, an unresolved core
cannot be ruled out even for this nearby cluster.  Even with the
repaired HST PSF, surface brightness data are unlikely to help in
clarifying the matter. The surface brightness profile is dominated by
the brightest stars and there are too few of them to give a good
statistical sample.  To get around this, more stars are required.  This
could be achieved by using star counts which go further down the main
sequence, but high resolution data will be needed to overcome the
crowding in the core.  If the distorting effects of the colour gradient
can be ironed out (and the number of stars involved is few enough that
overall star counts should not be affected) then a thorough set of high
resolution star counts should unambiguously discriminate between the
core-collapse and maximum-expansion phases and set firm bounds on the
core radius of NGC 6397.

There are two main points to consider in extending this comparison to
other globular clusters. On the positive side, more populous clusters
have more stars. This reduces the noise  associated with sampling. On
the negative side, all the other clusters are further away from us than
is NGC 6397, most considerably further.  The greater distance and the
higher crowding associated with higher densities increase the
resolution requirements for resolving the core and limits how far down
the main sequence we can count stars.  M15 is 2.7 magnitudes brighter
than NGC 6397, and all other things being equal, this corresponds to a
improvement in the signal to noise of the counting by a factor of 3.5.
On the other hand, it is 4.4 times further from us than NGC 6397,
giving an expected core radius of order 1\arcs, and the extra gain in
sampling is lost in the relatively larger area of the cluster we are
required to average over. This core radius is consistent with the limit
set by Yanny \et (1993). With a repaired HST the large wings of the
current point spread function will disappear and  the diffuse light
would once again be available for measuring the surface brightness
profile at high resolution. In this case discrimination on this basis
may be possible for M15.  What is required to definitively resolve
these very small cores is a combination of a large number of stars in
the core of the cluster of interest and high resolution. The defect of
surface brightness profiles is that they are hostage to the brightest
stars, which are proportionately few in number. Much better in terms of
counting statistics would be extensive star counts probing the main
sequence. Even higher resolution observations are then needed due to
the high degree of crowding. And even with the counts in hand, there
may just be too few stars in any given cluster. These requirements
highlight the difficulty of observing post-collapse cores in all but
the nearest clusters.

\section*{Acknowledgments}

\tx I wish to express my thanks to H. Cohn for suggesting this comparison
and to the referee, who's questions led to the clarification of several
previously obscure points.

\section*{References}

\bibitem Alcaino, G., Buonanno, R., Caloi, V., Castellani, V., Corsi, C. E.,
Iannicola, G.,  Liller, W. 1987, AJ, 94, 917  
\bibitem Bergbusch, P.A. \& VandenBerg, D.A. 1992, ApJS, 81, 163
\bibitem Chernoff, D.F.,  Weinberg, M.D. 1990, ApJ, 351, 121  
\bibitem Cohn, H.  1980, ApJ, 242, 765
\bibitem Djorgovski, S., Piotto, G., Phinney, E.S., Chernoff, D.F 1991 ApJ,
372, L41  
\bibitem Drukier, G.A. 1993, in preparation  
\bibitem Drukier, G.A., Fahlman, G.G.,  Richer, H.B. 1992, ApJ 386, 106  
\bibitem Drukier, G.A., Fahlman, G.G., Richer, H.B., Searle, L.,  Thompson,
I.B. 1993, in preparation 
\bibitem Fahlman, G.G., Richer, H.B., Searle, L.,  Thompson, I.B. 1989, ApJ,
343, L49  
\bibitem Gao, B., Goodman, J., Cohn, H., Murphy, B. 1991, ApJ, 370, 567  
\bibitem Grabhorn, R.P., Cohn, H.N., Lugger, P.M., Murphy, B.W. 1992, ApJ, 
392, 86  
\bibitem Grabhorn, R.P. \et, 1993, in Djorgovski, S., Meylan, G., eds.,
Dynamics of Globular Clusters, (ASP Conf. Ser.) in press  
\bibitem King, I.  1966, AJ, 71, 64  
\bibitem Lauer, T.R. \et 1991, ApJ, 369, L45 
\bibitem Lauzeral, C., Ortolani, S., Auri\`ere, M., Melnick, J. 1992, A\&A,
262, 63 (LOAM)  
\bibitem Lee, H.M., 1987, ApJ, 319, 772  
\bibitem Lee, H.M. \& Ostriker, J.P. 1987, ApJ, 322, 123
\bibitem Murphy, B.W., Cohn, H.N., Hut, P. 1990, MNRAS, 245, 335 
\bibitem Statler, T.S., Ostriker, J.P., Cohn, H.N. 1997, ApJ, 316, 626
\bibitem VandenBerg, D.A., Bell, R.A. 1985, ApJS, 58, 561  
\bibitem Yanny, B., Guhathakurta, P., Schneider, D.P., Bahcall, J. N. 1993,
PASP, in press

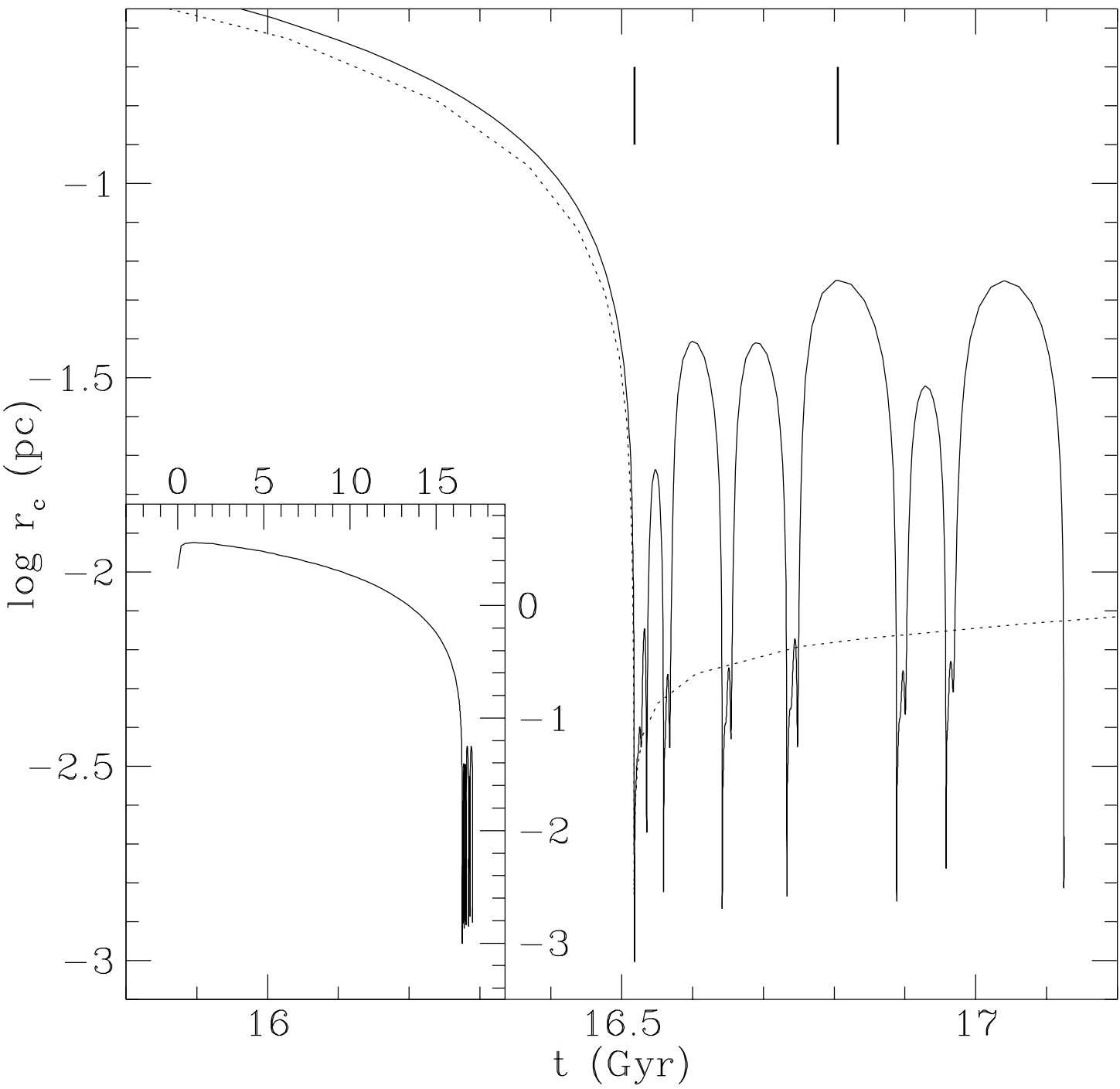
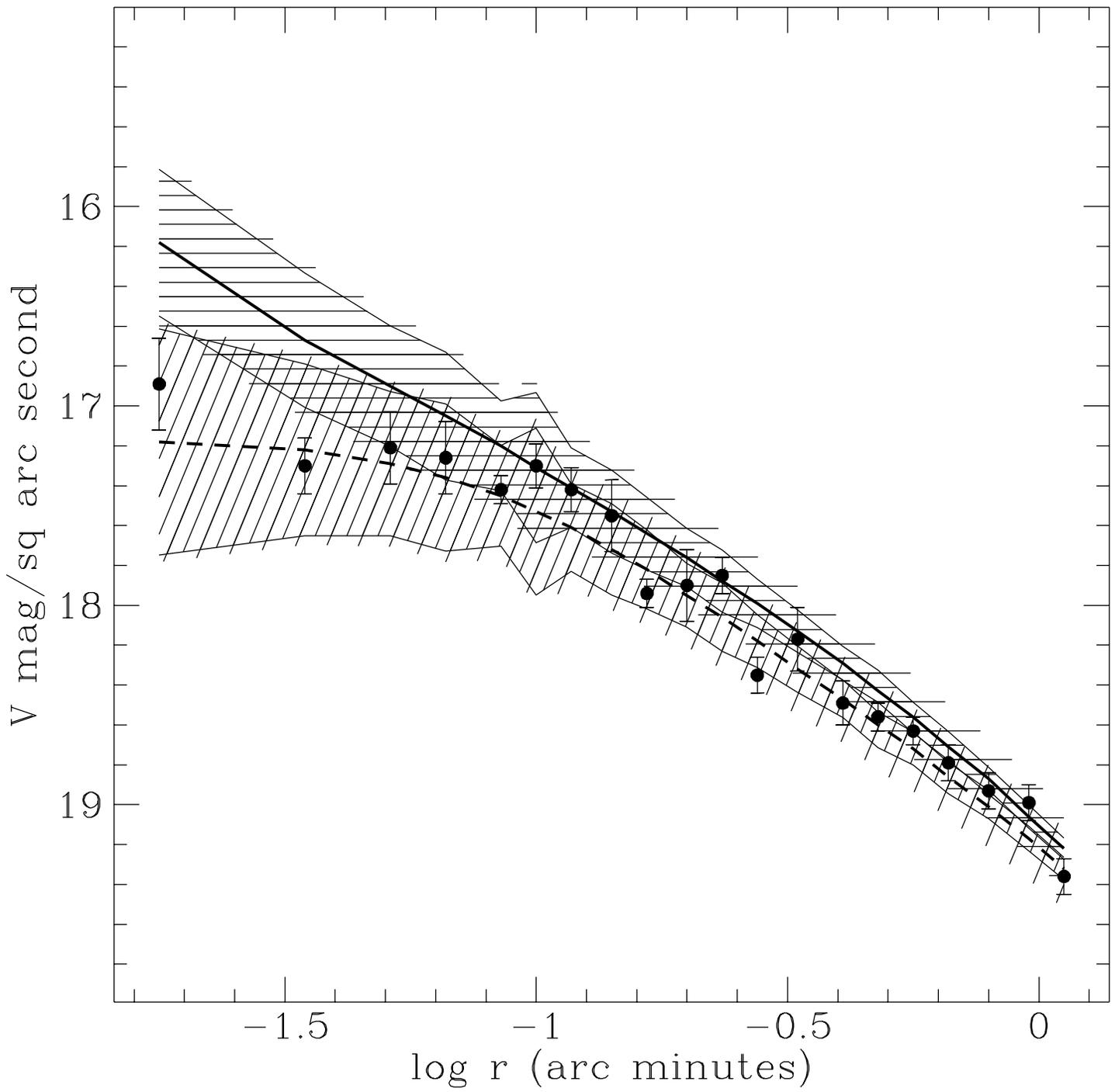
\bye